# Breaking BEC: Quantum evolution of unstable condensates

A. Kovtun[*] and M. Zantedeschi[†]

*Max-Planck-Institut für Physik, Föhringer Ring 6, D-80805 Munich, Germany
and Arnold Sommerfeld Center, Ludwig-Maximilians-Universität,
Theresienstraße 37, 80333 Munich, Germany*

In this work we numerically explore the quantum behavior of a classically unstable relativistic Bose-Einstein condensate (BEC). The main goal is to study the phenomenon of so-called quantum break time which amounts to a significant departure from a semiclassical mean-field description. It has been suggested previously that the existence of Lyapunov instability is crucial for a fast quantum breaking, chaos and scrambling. In order to clarify the issue, we work within the 2-PI effective action formalism and introduce a simple and very widely applicable dynamical criterion for identifying the timescale of quantum breaking. We indeed observe that the fast quantum break time is controlled by the Lyapunov exponent of the unstable BEC.

## I. INTRODUCTION

Quantum breaking is the idea that a given semiclassical solution, due to its quantum nature, might not be eternally faithful in describing the evolution of a system. In fact, due to quantum effects, a system might deviate in time from the semiclassical trajectory and significantly change its structure, therefore making the aforementioned mean-field approximation eventually unreliable. The concept that macroscopic objects can lose their classicality and become more "quantum" after a certain critical timescale was first introduced and developed in a series of papers [1–4] motivated by a microscopic composite picture of a black hole, and it was later generalized to various systems, such as inflationary cosmologies and axion field [5,6]. The outcome of these studies was that certain macroscopic systems, that are usually assumed to be well-described classically, in reality, exhibit a rather short quantum breaktime. There is a class of systems that seems to exhibit breaking much faster than the others.

Namely, in [7] the connection between phenomena of quantum breaking and chaos was established and it was argued that a many-body macroscopic system can undergo a maximally fast quantum breaking and become chaotic, provided it possesses a Lyapunov exponent $\gamma$, with the following formula for quantum break-time:

$$t_{qb} \sim \gamma^{-1} \log N, \qquad (1)$$

where $N$ is a certain macroscopic particle number (e.g., the number of off-shell gravitons in the black-hole case, or the particle number in a nonrelativistic BEC). In [7], this equation was explicitly checked on an example of a $1+1$ dimensional system with Lyapunov exponent, namely a nonrelativistic unstable BEC. The above mentioned quantity was derived by means of entanglement arguments. Moreover it was also suggested that quantum breaking and chaos represent the microscopic mechanisms behind the so-called phenomenon of quantum information scrambling and that the existence of Lyapunov exponent is crucial for a system to saturate the logarithmic bound on the fast scrambling time proposed in [8]. It is argued that these kind of systems are the ones which break the fastest.[1]

The results of [7] leave certain questions open. For example, it is unclear whether relativistic corrections would affect the above predicted timescale.

The purpose of this article is therefore to study these relativistic effects and to employ an alternative method allowing to compute the quantum-corrected evolution of the semiclassical solution initially assembled as a coherent state within mean-field approximation. Due to the absence of particle conservation, it is rather natural to expect a similar behavior under the replacement $N \to Q$ which is the conserved charge of the system in the relativistic case. This replacement is anyway absolutely not obvious, as the relativistic theory possesses a much wider spectrum. In this work, a law similar to (1) was found for a relativistic $1+1$-dimensional model endowed with $U(1)$ symmetry, attractive self-interaction and periodic boundary conditions. Namely, we verified that the quantum break time is given by

$$t_{qb} = \gamma^{-1} \log Q + \text{constant},  \qquad (2)$$

where $Q$ is the total dimensionless charge of the configuration,[2] $\gamma$ is the Lyapunov exponent associated with the system and the small constant presence will be shown to be related to the chosen criterion used to extract $t_{qb}$. Moreover, note how this timescale happens to be infinite in the semiclassical limit (namely as $\hbar \to 0$, $t_{qb} \to \infty$), therefore ensuring the captured effect to be genuinely quantum and not visible at the classical level.

---

[*]akovtun@mpp.mpg.de
[†]michaelz@mpp.mpg.de

[1]There are systems where $t_{qb} \propto \hbar^{-1}$ c.f. [1], which is much longer than (1).

To obtain it we work in the semiclassical framework of 2PI effective action. In fact, with this method we are able to study the unitary Minkowski-time evolution of a quantum coherent state mimicking a relativistic Bose-Einstein condensate. In turn, this allows us to check how the system dynamically deviates from the mean-field condensate solution. To do so, we introduce a criterion which can be easily applied for identifying the timescale associated to quantum breaking for various semiclassical systems well-described within mean-field approach. Namely, we look at the dynamical evolution of the conserved integral quantity constraining the system, i.e., charge.[3] Indeed, as discussed below, such quantity is exactly conserved due to Noether's theorem. However, since we are working within the 2PI framework, such composite quantity receives two contributions from the one- and the two-point expectation values respectively (note that these quantities balance each other out because of the above mentioned symmetry to each loop order in case of an $\hbar$ expansion [9]). We will refer to the former as classical charge $Q_{cl}$ and to the latter as quantum charge $Q_q$. At the beginning of the evolution, the ratio between $Q_{cl}$ and $Q_q$ is fixed by the initial conditions. As one would expect, for a coherent state describing a classical configuration we have that $Q_{cl} \gg Q_q$. However, as the system evolves, this needs not be the case anymore. It is therefore natural to define quantum breaking as the timescale when the two above-mentioned quantities become comparable. A great advantage from this criterion is obtained, as no need to deal with rescattering effects at the microscopic level is necessary, although one could infer it from the diagrams retained within a given 2PI expansion scheme. It is worth mentioning that in principle different integral quantities could serve as a mean to estimate the quantum break time as long as their conservation constrains the dynamics of the system. For example, one could equally use energy or, in the case of a nonrelativistic system, occupation number. In the present article, however, our main focus is charge since, as we will see below, the contributions from the 1- and the 2-point expectation values split in a very clear and manageable manner.

In [10] we have already applied this approach to study the relativistic BEC with repulsive self-interaction and analyzed the dependence of quantum break-time on the charge of the initial coherent state. Although we did not extract analytical dependence, it is apparent that for extremely big and extremely small charges, quantum break-time is at least polynomial in terms of charge, and becomes non-negligible on very long timescales. We would like to draw the reader's attention to the observation that in case of cosmic axion [6] the quantum break time is also polynomial in the number of constituents initially accumulated in the condensate, although in that case, it is derived by means of perturbative calculations. No conflict between our and their result was found. We therefore conclude that our newly introduced simple criterion is very efficient in determining quantum breaking, as not only it is applicable to a wide class of systems, but it also agrees with the quantum breaking timescale derived with different methods in the literature (based on a perturbative calculation in [6] and a nonperturbative one in [7]).

Before we explain the method and the system, we will first discuss the relation between quantum breaking and scrambling. After that we will explain the method and criterion used to extract quantum break-time and present the results.

## II. VARIOUS APPROACHES TO QUANTUM BREAKING

There are plenty of systems that are dominantly classical and therefore they can be studied within semiclassical methods, e.g., mean field. Classicality of a physical object implies that there is no need to identify its internal degrees of freedom in order to predict its evolution as a whole, because quantum interactions are subleading when compared to the classical dynamics. As long as the interaction between constituents, namely the collective coupling, is much smaller than unity, quantum breaking matters on very long timescales. *This is why we do not observe quantum breaking in everyday life.* Indeed, with non-negligible collective coupling (in which the system is said to be critical), quantum dynamic might lead to substantial deviations from the classical description. Intuitively, the picture of quantum breaking is simple: it is due to critical or close-to-critical interacting quantum constituents, which with time invalidate the tree-level description as the system becomes more and more quantum. When this happens, the system is said to have undergone quantum breaking. As pointed out earlier there is a huge amount of very different systems, hence a natural question to ask is whether there is a way to explain quantum breaking generically, regardless of the particular details of a given system.

---

[2]Mean-field expectation value of the dimensionless operator of charge (meaning that $\hbar$ is kept and not set to 1).
[3]We would like to remark that one could choose any other conserved integral quantity provided that it is possible to split it in a certain way, which we will later name as *quantum* and *classical* contributions.

One approach developed to understand quantum breaking is deeply intertwined with the so-called scrambling phenomenon, which is considered to be the microscopic mechanism leading a system toward quantum breaking. By scrambling here it is meant the process of thermalization described by the reduced density matrix (defined on a subset of the Hilbert space) which leads to the scrambling of information and growth of entropy. The timescale associated to this process is called scrambling time and in [7] this timescale was associated with quantum breaking as well. In fact, in Ref. [7], the initial state is chosen to be macroscopically occupied with zero-momentum modes and the subsequent evolution leads to a significant smearing of this distribution. As one can see, this definition is purely informational and the entropy growth can be computed by evaluating the unitary evolution of the reduced density matrix. However, one must take into account that systems subjected to scrambling may not be initially macroscopic or coherently assembled (namely "classical like"). Therefore, the so-called scrambling phenomenon does not always imply actual decoherence or a substantial deviation from the semiclassical trajectory. More precisely it stands for thermalization of the reduced density matrix.[4] This means that scrambling does not always implies quantum breaking. However the opposite is true. Looking at quantum breaking as the process leading to significant deviation from an initial macroscopic state, we can say that quantum breaking has occurred when information about such initial state is substantially lost. This can be reflected in the observation that the notion of initial degrees of freedom or information about the way they were assembled within the initial state is lost (or dramatically long time is needed for this information to be restored). In other words we can say that if quantum breaking takes place, the system necessarily scrambled, i.e.,

$$\text{quantum breaking} \subset \text{scrambling}.$$

A more elaborate discussion of the relation between quantum breaking and scrambling can be found in [7].

However, it is quite apparent that the estimation of the scrambling time poses significant difficulties, as it requires the knowledge of the object microscopic structure, which is not always accessible. Indeed resolving the inner structure in terms of off-shell degrees of freedom for most of the known semiclassical systems is not possible. It is instructive to have a different method to estimate quantum braking. Namely, we can think of quantum breaking as a cascade of re-scattering processes between degrees of freedom assembled in the initial macroscopic state. Hence, as the system evolves, the rescattering of this quanta leads to a different quantum state. Eventually, if the result of this tower of S-matrix processes is far enough from the classical description we can conclude that quantum breaking took place. Therefore, if one encodes some information in the in-state, as a result of scattering events, such information will be encoded differently in the final state (for example different momentum-modes are occupied). It may even occur that an exponentially long time is needed to recover the initially encoded information. This is exactly a consequence of scrambling.[5]

## III. 2PI EFFECTIVE ACTION APPROACH TO QUANTUM BREAKING

### A. General definitions

We are interested in studying the nontrivial time evolution of a Bose-Einstein-condensate (BEC) in $1+1$ dimensions. For this purpose, the tools employed in nonequilibrium field theory are very advantageous: more specifically, we are going to use the effective action for composite operators. This framework has already proved its efficiency when applied to nonequilibrium configuration in statistical field theory [11,12].

The formalism of the 2PI effective action was first introduced in [13,14] and an efficient computational procedure was developed in [15]. Here we recap the main points to establish our notation.

Let us define the generating functional of the theory as

$$Z[J,K] = \int \mathcal{D}\varphi_a \exp\left(i\left(S[\varphi_a] + \int \varphi_a(x) J_a(x) + \frac{1}{2} \int \int \varphi_a(x) K_{ab}(x,y) \varphi_b(y)\right)\right). \quad (3)$$

Here, $J_a(x)$ and $K_{ab}(x,y)$ are the 1- and 2-particle sources and by $x$ we mean the set of all space-time coordinates.

The corresponding generating functional for connected diagrams is

$$W[J,K] = -i \ln Z[J,K]. \quad (4)$$

To derive the 2PI effective action we perform a Legendre transform with respect to both sources

---

[4]We remark this because unitary evolution may not lead to significant deviations from the initial state, even if the latter is macroscopic and is not a pure eigenstate of the Bose Hamiltonian.

[5]For instance, in the cosmic axion case [6] this is the precise microscopic mechanism behind quantum breaking, which shows its deep relation to scrambling. In particular, for this simple system, it was possible to extract the quantum breaking time by means of perturbative estimation: in fact, a microscopic perturbative resolution of the internal degrees of freedom was possible.

$$\Gamma[\phi, G] = W[J, K] - \int \frac{\delta W[J, K]}{\delta J_a(x)} J_a(x)$$
$$- \frac{1}{2} \int \int \frac{\delta W[J, K]}{\delta K_{ab}(x, y)} K_{ab}(x, y), \quad (5)$$

where we introduced

$$\begin{cases} \frac{\delta W[J,K]}{\delta J_a(x)} = \phi_a(x), \\ \frac{\delta W[J,K]}{\delta K_{ab}(x,y)} = \frac{1}{2}(\phi_a(x)\phi_b(y) + G_{ab}(x, y)), \end{cases} \quad (6)$$

with $\phi_a(x)$ the expectation value of the field, and $G_{ab}(x, y)$ the full connected propagator of the theory.

In the limit of vanishing sources, one obtains the stationary conditions for the effective action

$$\begin{cases} \frac{\delta \Gamma[\phi,G]}{\delta J_a(x)} = 0, \\ \frac{\delta \Gamma[\phi,G]}{\delta G_{ab}(x,y)} = 0. \end{cases} \quad (7)$$

Solving these stationary conditions, accompanied with appropriate renormalization conditions, one can eventually compute the effective action or other quantities related to its functional derivatives.

We would like to emphasize that in the limit of vanishing sources variables $\phi_a(x)$ and $G_{ab}(x, y)$ become exact 1- and 2-point correlation functions. Due to this essential feature of the formalism one can dynamically account for the backreaction of quantum fluctuations on the 1-point expectation value and vice versa. In fact, within this framework both the propagator $G_{ab}(x, y)$ and the mean-field $\phi_a(x)$ are endowed with their own dynamics and interact between each other. Within mean-field treatment $\phi_a(x)$ is considered as an expectation value of coherently excited modes, approximating the BEC. Therefore, aforementioned features of the formalism allow us to study the dynamical decoherence happening under the influence of quantum fluctuations encoded in the 2-point correlator. This is one of the greatest advantages of the 2PI effective action formalism.

It is possible to explicitly compute the 2PI effective action as a loop expansion, which leads to [15]

$$\Gamma[\phi, G] = S[\phi] + \frac{i}{2}\operatorname{tr}\ln G^{-1} + \frac{i}{2}\operatorname{tr}(G_0^{-1}G) + \Gamma_2[\phi, G], \quad (8)$$

where $\Gamma_2[\phi, G]$ is the sum of the vacuum-to-vacuum 2-particle irreducible diagrams with $n$ – loops, $n \geq 2$, computed with the following Feynman rules:
 (i) Every internal line carries the propagator $G_{ab}(x, y)$,
 (ii) Vertices are given by the nonlinear part of the shifted action $S[\varphi_a + \phi_a]$,

and

$$G_{0,ab}^{-1}(x, y) = -i \frac{\delta^2 S[\phi]}{\delta \phi_a(x) \delta \phi_b(y)}. \quad (9)$$

Summarizing all the definitions and imposing stationary conditions of the 2PI functional, the dynamical equations of motion for the 1 and 2-point Green's function can be obtained as

$$\begin{cases} \frac{\delta S[\phi]}{\delta \phi_a(x)} + \frac{i}{2}\operatorname{tr}\left(\frac{\delta G_0^{-1}}{\delta \phi_a(x)} G\right) + \frac{\delta \Gamma_2[\phi,G]}{\delta \phi_a(x)} = 0, \\ G_{ab}^{-1}(x, y) - G_{0ab}^{-1}(x, y) + 2i \frac{\delta \Gamma_2[\phi,G]}{\delta G_{ab}(x,y)} = 0. \end{cases} \quad (10)$$

Note that to solve these equations a proper time-integration contour must be specified. This is the focus of next paragraph.

### B. *In-in* contour

As mentioned in the introduction, in order to obtain dynamical equations of motion that are causal, we choose the Schwinger-Keldysh time contour. Since the dynamics depends only on the past, we can easily simulate the difficult integral differential equations (10) using a finite difference scheme. It is simpler to evaluate the diagrams using the approach described in [16]. Namely, we decompose the connected Green's function in two parts, $F$ and $\rho$, defined as

$$G_{ab}(x, y) = F_{ab}(x, y) - \frac{i}{2} \operatorname{sgn}_\mathcal{C}(x^0 - y^0) \rho_{ab}(x, y), \quad (11)$$

$$F_{ab}(x, y) = \frac{1}{2}\langle\{\phi_a(x), \phi_b(y)\}\rangle, \quad (12)$$

$$\rho_{ab}(x, y) = i\langle[\phi_a(x), \phi_b(y)]\rangle, \quad (13)$$

where the $\operatorname{sgn}_\mathcal{C}(x^0 - y^0)$ is taken along the *in-in* time contour and guarantees proper time ordering.[6] Here, $F_{ab}(x, y)$ is known as statistical propagator, while $\rho_{ab}(x, y)$ is related to the spectrum of the theory.

Substituting this decomposition in the second equation of (10) two equations for $F_{ab}(x, y)$ and $\rho_{ab}(x, y)$ are obtained. Thus, we have a set of three second-order differential-integral equations (29) and three unknown functions (omitting counting with respect to field indices) $\phi_a, F_{ab}$, and $\rho_{ab}$. As already mentioned, to solve this system we have to complement it with a set of appropriate initial and boundary conditions. This will be done in Sec. III.

---

[6]For the Feynman propagator one can see that it will be just ordinary sign function $\operatorname{sgn}(x^0 - y^0)$.

### C. Recipe for deriving the quantum break time

The goal of this work is to study the dynamical decoherence of a classically unstable BEC. Within the 2PI approach, both correlators $\phi_a(x)$ and $G_{ab}(x,y)$ have their own mutually-interacting dynamics. It is natural to ask what is a good indicator to determine whether a significant departure from the classical trajectory took place, namely, how to understand when decoherence, which is associated with quantum breaking, happened. Since the initial classical-like configuration consists of coherently excited zero-momentum modes, we define quantum-breaking as the moment after which most of these quanta no longer contribute to the 1-point expectation value $\phi_a(x)$. This is due to rescattering effects that, with time, lead to the occupation of nonzero momentum modes. The latter quantity is encoded in the propagator $G_{ab}(x,y)$. Now we will give a general recipe and then implement it to our particular problem.

#### 1. Recipe for arbitrary system

Equipped with the necessary knowledge about 2-PI effective action supplemented with nonequilibrium Schwinger-Keldysh time-integration contour we are finally ready to formulate the quantum breaking criterion. In the following a general approach for studying the dynamical quantum decoherence of a classical-like system is given.

It is well known that classical solutions are commonly described within background-field treatment. Namely, a coherent field configuration is described in terms of mean-field expectation value $\phi_a(x)$, solving classical equations of motion. Using saddle-point approximation or ordinary 1-PI effective action one can compute quantum corrections and Green's function in the background of this configuration. However, within these approaches quantum fluctuations are just functions of a given stationary background. On the opposite side, within 2-PI treatment, quantum fluctuations, which are encoded in the exact propagator $G_{ab}(x,y)$, follow their own interacting evolution, which is of course intertwined with the evolution of the expectation value $\phi_a(x)$ according to Eq. (10). Therefore both quantities mutually interact and simultaneously evolve. Our criterion for quantum breaking is based on the validity of the semiclassical approximation, implying the existence of a quantity $\mu$ (e.g., charge or particle number) satisfying following conditions:

(1) It is conserved along the dynamical flow (deduced via Nöther identities)

$$\frac{d\mu}{dt} = 0$$

(2) Reflects classicality of the system

$$\mu \xrightarrow{\hbar \to 0} \text{constant},$$

*Important remark*: We should clarify this particular point by giving few examples.

Let us take the energy as an example of classical quantity. The energy of a classical system survives in the $\hbar \to 0$ limit as it can be measured with a classical probe, hence, the term "classical." For charge or particle number the story is different. To illustrate their relation with $\hbar$ we recall that the angular momentum in 2-dimensional systems is quantized in $\hbar$-step, namely, $L = \hbar l$, where $l \in \mathbb{Z}$. One can see that it is directly analogous to the $U(1)$-charge considered in the current work. The quantity $L$ can be classically large and remain finite. In the semiclassical treatment, however, the dimensionless quantity $l$ is expressed in terms of mean-coordinate (the analog of the mean-field for field theory) and the mean-coordinate expression for this dimensionless quantity scales as $\langle \hat{l} \rangle \sim 1/\hbar$. Hence, the dimensionful quantity $L$ survives in the classical limit. By analogy the dimensionless operator $\hat{Q}$ corresponds to the $U(1)$-charge having quantized eigenvalues. The mean-field value of $\langle \hat{Q} \rangle$ scales as $\langle \hat{Q} \rangle \sim 1/\hbar$, while in the classical limit it is the quantity $\hbar \langle \hat{Q} \rangle$ that survives. This quantity does not depend on $\hbar$ and is continuous in the classical limit as it expresses some continuous conserved quantity.[7] To some up $Q_{cl} = \hbar \langle \hat{Q} \rangle$, where $\langle \hat{Q} \rangle = Q$ is the mean-field expression, which we will keep calling classical because throughout the work we will use $\hbar = 1$ units.

(3) Since we expanded $\Gamma[\phi, G]$ in powers of $\hbar$, quantity $\mu$ expressed in terms of the effective fields $\phi$ and $G$ can be expanded semiclassically, i.e.,

$$\mu = \mu_{cl}(t) + \mu_q(t), \qquad (14)$$

where both contributions can change with time, but whose sum is conserved. We will call them according to subscripts "classical" and "quantum" because throughout the rest of the work we employ units $\hbar = 1$. In the following we will employ terminology which regards quantities expressed in terms of mean-field as classical.

(4) At the beginning of the evolution must be satisfied

$$\mu_{cl}(0) \gg \mu_q(0). \qquad (15)$$

Thus, if $\hbar$ is restored classical limit gives

$$\frac{\mu_q(0)}{\mu_{cl}(0)} \xrightarrow{\hbar \to 0} 0. \qquad (16)$$

---

[7]We will expand on that later on.

Hence, we define the quantum break time $t_{qb}$ as the moment when

$$\mu_{cl}(t_{qb}) \simeq \mu_q(t_{qb}). \qquad (17)$$

In the following, we will present a concrete study based on (15) and (17).

### 2. Generalized Ward-Takahashi identities

According to [17], if the theory has a continuous global symmetry group $\mathcal{O}$ such that a field transformation $O \in \mathcal{O}$ leaves the Lagrangian invariant

$$\phi_a \to O_{ab}\phi_b, \qquad O_{ab} = \mathbb{1}_{ab} + i\epsilon^\alpha \tau^\alpha_{ab} + \mathcal{O}(\epsilon^2),$$
$$\mathcal{L}(\phi_a, \partial\phi_a) = \mathcal{L}(O_{ab}\phi_b, \partial O_{ab}\phi_b),$$

where $\tau^\alpha_{ab}$ are generators of the corresponding Lie algebra, then there is a set of integrals of motion given by

$$\int_\mathcal{C} dx \frac{\delta \Gamma[\phi, G]}{\delta \phi_a(x)} \tau^\alpha_{ab}\phi_b(x)$$
$$+ \int_\mathcal{C}\int_\mathcal{C} dx dy \frac{\delta \Gamma[\phi, G]}{\delta G_{ab}(x,y)}(\tau^\alpha_{ac}\delta_{bd} + \tau^\alpha_{ad}\delta_{bc})G_{cd}(x,y) = 0. \qquad (18)$$

Integrating by parts these identities one can extract the conserved charges. In the case of $SO(2)$ symmetry, there is only one generator and, correspondingly, a single conserved current whose charge is given by

$$Q = \int d\mathbf{x} \lim_{y \to x} \epsilon_{ab}\partial_{x^0}(\phi_a(x)\phi_b(y) + G_{ab}(x,y)), \qquad (19)$$

upon periodic boundary conditions, which we employ in order to preserve space-translation invariance. In this equation $\epsilon_{ab}$ is the 2-dimensional Levi-Civita tensor and $d\mathbf{x}$ denotes integration over spatial coordinates. In terms of decomposition (11), the charge is

$$Q = \int d\mathbf{x} \lim_{y \to x} \epsilon_{ab}\partial_{x^0}(\phi_a(x)\phi_b(y) + F_{ab}(x,y)). \qquad (20)$$

This quantity is conserved along the evolution of the system regardless of the given approximation. This not only proved to be a very useful criterion to verify the reliability and the stability of simulations, but it also allowed to establish a quantum-breaking criterion.

### 3. Quantum-breaking criterion based on Q

Using the results of previous subsections we can now easily suggest a very useful way to identify the timescale associated to quantum breaking. It should be noted that charge (19) is conserved at any finite $\mathcal{O}(\hbar)$ approximation as shown in [17]. Moreover, it splits nicely in terms of 2-PI effective action variables $\phi_a(x)$ and $G_{ab}(x,y)$ as it can be seen from (19), or more precisely $\phi_a(x)$ and $F_{ab}(x,y)$, as shown in (20). Therefore, the charge expressed in terms of the mean-field $\phi_a(x)$ is

$$Q_{cl}(x^0) = \int_0^L dx^1 \lim_{y \to x} \epsilon_{ab}\partial_{x^0}\phi_a(x)\phi_b(y)$$
$$= \int_0^L dx^1 (\dot{\phi}_1(x)\phi_2(x) - \dot{\phi}_2(x)\phi_1(x)), \qquad (21)$$

where "cl" stands for *classical*.[8] This quantity is given by the mean-field only.

Restoring $\hbar$ the above expression becomes

$$Q_{cl}(x^0) = \int_0^L dx^1 \hbar(\dot{\phi}_1(x)\phi_2(x) - \dot{\phi}_2(x)\phi_1(x))$$
$$= \hbar\langle \hat{Q} \rangle, \qquad (22)$$

showing that the mean-field average $\hat{Q}$ diverges in the $\hbar \to 0$ limit.

The second contribution is

$$Q_q(x^0) = \int_0^L dx^1 \lim_{y \to x} \epsilon_{ab}\partial_{x^0} F_{ab}(x,y), \qquad (23)$$

which accounts for the amount of charge accumulated in the connected 2-point function where "q" stands for *quantum*. It is fundamental to underline the fact that such a splitting between classical and quantum contribution is natural. This is encoded in the high classicality of the chosen initial configuration (corresponding to the saddle point, see Sec. IV D) i.e.,

$$Q_{cl}(t=0) \gg Q_q(t=0). \qquad (24)$$

From this, if one restores $\hbar$ explicitly, it follows that

$$\frac{Q_q(0)}{Q_{cl}(0)} \overset{\hbar \to 0}{\to} 0, \qquad (25)$$

and therefore, in the semiclassical limit $\hbar \to 0$, the standard stationary background field picture is recovered as it was demanded in (16). Both quantities are scalar functions of time and evolve according to Eq. (10), but their sum is conserved due to Ward-Takahashi identities. However, during evolution both charges will change their values. Correspondingly, nonzero momentum modes are populated, and the classical-like field undergoes decoherence. Therefore we look at how the ratio $Q_q(t)/Q_{cl}(t)$ changes in time. This will serve as a mean to measure the departure from the initial configuration. In fact, it is natural to define the quantum break time as the moment when

$$Q_q(t_{qb}) \simeq Q_{cl}(t_{qb}), \qquad (26)$$

as it was stated in (17) and therefore we use (26) as a way to estimate the quantum break time for the given system.

---

[8]Once again we regard term "classical" to denote the mean-field contribution.

It should be noted that in the theory endowed with $SO(2)$ symmetry there is a global conserved charge, which is a bilinear function of field operators. This is a key factor from which the emergence of such a clear decomposition of the full charge (20) in classical (21) and quantum (23) components follows. However, if the system is not endowed with such a conserved quantity, a different quantity should be used. In fact, every fundamental system is by default endowed with integrals of motion connected to the Poincaré group. Therefore, another possibility would be to use the full energy functional, which can also be split into quantum and classical contributions: the classical part stands for the classical energy, namely an expression in terms of the 1-point correlation function and the remaining part is quantum since it depends on the 2-point function. In this case, however, the quantum part is a function of both $G(x, y)$ and $\phi(x)$.

### D. Approximation of the 2PI effective action

Before moving forward it is worth explaining the limits of the approximations employed in the 2PI effective methods.

As usual we employ conventional expansion in the number of loops developed in [15]. For our purpose it is sufficient to expand $\Gamma_2(\phi, G)$ [see (8)] up to the second order in loops. This corresponds to the inclusion of the following diagrams:

$$\Gamma_2(\phi, G) = \underset{\lambda}{\infty} + \lambda\phi_a \underset{}{\bigcirc} \lambda\phi_b + \mathcal{O}(\hbar^3), \qquad (27)$$

where $\lambda$ is the quartic coupling. In terms of effective action variables we can rewrite this expression in the form

$$\Gamma_2[\phi, G] = -\frac{\lambda}{16} \int_\mathcal{C} dx (2G_{ab}(x,x)G_{ab}(x,x) + (G_{aa}(x,x))^2) + \frac{i\lambda^2}{16} \int_\mathcal{C} dx \int_\mathcal{C} dy (G_{ab}(x,y)\phi_a(x)\phi_b(y)G_{cd}(x,y)G_{cd}(x,y)$$
$$+ 2\phi_a(x)G_{ab}(x,y)G_{bc}(x,y)G_{cd}(x,y)\phi_d(y)) + \mathcal{O}(\hbar^3), \qquad (28)$$

where $\mathcal{C}$ denotes the time integration contour (for a derivation of this expression see Appendix in [10]).

After substituting this expression for $\Gamma_2^{(2)}[\phi, G]$ in stationary conditions (10) and making use of decomposition (11) we can derive the equations of motion for both parts of the Green's functions and the mean-field

$$\begin{cases} -\frac{\delta S[\phi]}{\delta \phi_a(x)} + \frac{\lambda}{4} F_{cc}(x,x)\phi_a(x) + \frac{\lambda}{2} F_{ab}(x,x)\phi_b(x) &= \int_0^{x^0} dy^0 \int_0^L dy^1 \Sigma^\phi_{ab}(x,y)\phi_b(y), \\ (\partial^2 \delta_{ab} + M^2_{ab}(x))F_{bc}(x,y) &= \int_0^{y^0} dz^0 \int_0^L dz^1 \Sigma^F_{ab}(x,z)\rho_{bc}(z,y) - \int_0^{x^0} dz^0 \int_0^L dz^1 \Sigma^\rho_{ab}(x,z)F_{bc}(z,y) \\ (\partial^2 \delta_{ab} + M^2_{ab}(x))\rho_{bc}(x,y) &= -\int_{y^0}^{x^0} dz^0 \int_0^L dz^1 \Sigma^\rho_{ab}(x,z)\rho_{bc}(z,y), \end{cases} \qquad (29)$$

where the mass matrix $M_{ab}(x)$ is the local part of the self-energy defined as

$$M^2_{ab}(x) = (m^2 + \delta m^2)\delta_{ab} + \frac{\lambda}{4}(\phi_c^2(x) + F_{cc}(x,x))\delta_{ab}$$
$$+ \frac{\lambda}{2}(\phi_a(x)\phi_b(x) + F_{ab}(x,x)). \qquad (30)$$

The functions $\Sigma^\phi_{ab}(x,y)$, $\Sigma^\rho_{ab}(x,y)$ and $\Sigma^F_{ab}(x,y)$ are explicitly derived in the Appendix of [10], where we have also presented a detailed discussion regarding the relevance of the used diagrams for the field dynamics. Here we will just outline the most important statements.

The contribution given by the first diagram in (27) is local, and therefore a trivial correction to the e.o.m. The second diagram, however, is responsible for the interaction between $\phi$ and $G$. As long as we are in the perturbative regime and $t < t_{qb}$, it is the most relevant diagram for the dynamics and the one responsible for quantum breaking. After $t_{qb}$, interactions between nonzero momentum modes (i.e., those modes which do not comprise the background field) might become of comparable importance and, therefore, higher orders in the loop expansion should be taken into account. It is worth stressing, anyway, that the above discussion implies that the two-loop approximation is enough to reliably obtain the quantum break time via condition (26). Finally, note that the second diagrams scales schematically as $\lambda(\lambda\phi^2/m^2)$, which is exactly the necessary effective coupling that was used to capture $1/N$ effects in [5].

In order to solve equations (29) we have to specify initial and boundary conditions. This is simple for $\rho_{ab}(x, y)$, because its initial values are given by commutation relations [10,18]. But in order to get initial conditions for $\phi_a(x)$ and $F_{ab}(x, y)$ we have to consider the semiclassical theory of BEC, which will be addressed in the next section. Finally periodic boundary conditions are chosen.

## IV. CLASSICAL THEORY

In this section we introduce classical theory and analyze its spectrum.

Consider a real scalar field endowed with $SO(2)$ global symmetry with attractive quartic self-interaction in a $1+1$ dimensional finite box of size $L$. The action of this theory is:

$$S[\varphi_a] = \int_0^T dx^0 \int_0^L dx^1 \left( \frac{1}{2}(\partial_\mu \varphi_a)^2 - \frac{1}{2}m^2 \varphi_a^2 + \frac{\lambda}{16}(\varphi_a^2)^2 \right), \tag{31}$$

with $\lambda > 0$. In the following we will consider this action upon periodic boundary conditions for spatial variable of the field $\phi_a(x)$.

For future purposes we list here two integrals of motion which are relevant for our analysis. These are the classical energy

$$E^{(cl)} = \int_0^L dx^1 \left( \frac{1}{2}\left(\frac{\partial \varphi_a}{\partial t}\right)^2 + \left(\frac{\partial \varphi_a}{\partial x}\right)^2 + \frac{1}{2}m^2 \varphi_a^2 - \frac{\lambda}{16}(\varphi_a^2)^2 \right), \tag{32}$$

and the classical charge

$$Q^{(cl)} = \int_0^L dx^1 \, (\dot{\varphi}_1(x)\varphi_2(x) - \dot{\varphi}_2(x)\varphi_1(x)), \tag{33}$$

which is literally the same as the one given by (21), but while here the field is classical, in (21) $\phi_a(x)$ is the expectation value of the operator.

This theory enjoys plenty of classical stationary solutions. These solutions can be obtained by the following ansatz

$$\varphi_a(t, x) = \sqrt{2} R(\omega t)_{ab} \tilde{f}_b(x), \tag{34}$$

where $R(\theta)$ is the usual $SO(2)$ rotation matrix and $\omega$ is the integration constant parametrizing all possible solutions. Using the $SO(2)$ covariance of the equations we can choose $\tilde{f}_b(x) = (f(x), 0)^T$. Then, two equations of motion are reduced to a single one

$$\frac{d^2 f}{dx^2} + (\omega^2 - m^2)f + \frac{\lambda}{2}f^3 = 0. \tag{35}$$

We are going to look for all solutions satisfying periodic boundary conditions.

Two solutions to (35) can be found: namely a homogeneous one, the condensate, and a localized one, the bright soliton [19]. In the following their relation and stability are discussed.

### A. Condensate

The condensate solution is given by the homogeneous configuration

$$f(x) = \sqrt{\frac{2(m^2 - \omega^2)}{\lambda}}, \tag{36}$$

where the frequency covers the range $\omega \in (0, m)$. For this solution the integrals of motion (32) and (33) are

$$E_{b.c.} = \frac{L}{\lambda}(m^4 + 2m^2\omega^2 - 3\omega^4), \tag{37}$$

$$Q_{b.c.} = \frac{4\omega L(m^2 - \omega^2)}{\lambda}, \tag{38}$$

where subscript "b.c." stands for Bose condensate. Notice that for this attractive interaction the energy of the configuration is lower than the one of free particles, i.e., $E(\omega) \le mQ(\omega)$.

### B. Bright soliton

The second solution is the well-known bright soliton. In the nonrelativistic limit, the classical bright soliton was studied in [19,20]. Here we focus on the relativistic case.

The solution is showed for different $\omega$'s in Fig. 1 and is given by

$$f(x) = \frac{4K(\mu)}{L\sqrt{\lambda}} dn\left( 2K(\mu)\frac{x}{L} \bigg| \mu \right), \tag{39}$$

where $K(\mu)$ is the elliptic integral of the first kind, $dn(x|\mu)$ is the Jacobi elliptic function and $\mu$ is fixed by the condition

$$4K(\mu)^2(2 - \mu) = L^2(m^2 - \omega^2), \tag{40}$$

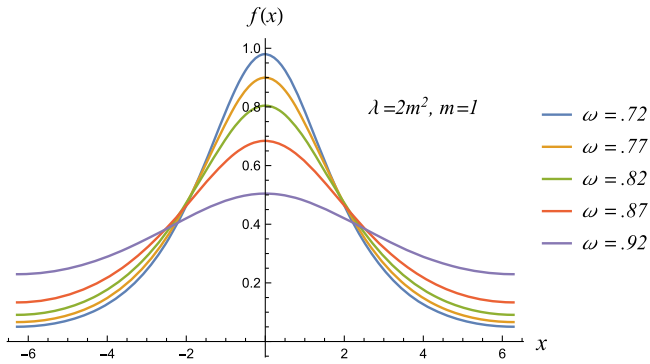

FIG. 1. Bright soliton solution for different frequencies $\omega$ and length $L = 4\pi m^{-1}$.

for which the solution exists if $\omega \in (0, \omega_{cr})$ and the critical frequency is given by

$$\omega_{cr} = \sqrt{m^2 - \frac{2\pi^2}{L^2}}. \quad (41)$$

The integrals of motion (32) and (33) corresponding to the bright soliton are

$$E_{b.s.}(\omega) = \frac{16(4(\mu-1)K(\mu)^4 + L^2(4\omega^2 + 2)K(\mu)E(\mu))}{3\lambda L^3}, \quad (42)$$

$$Q_{b.s.}(\omega) = \frac{32\omega}{L\lambda} E(\mu)K(\mu). \quad (43)$$

where subscript "*b.s.*" stands for bright soliton, $E(\mu)$ is the elliptic integral of the second kind and $\mu = \mu(\omega)$ is given by (40).

### C. Classical stability as the reflection of an interplay between two solutions

We see that both the homogeneous BEC and the inhomogeneous bright soliton solutions depend on a single integration constant $\omega$. This integration constant defines their charge and energy. Therefore we have several branches of solutions parametrized by $\omega$. Let us focus first on the blue branch in Fig. 2, representing the condensate solution for different values of $\omega$ and correspondingly for different charges. We note that $\omega$ varies from 0 to $m$, where values closer to $m$ have smaller charges. The orange branch instead represents the bright soliton configurations with different $\omega$. For the bright soliton we have $\omega \in (0, \omega_{cr})$, from which it follows that for $\omega \in [\omega_{cr}, m]$ the homogeneous solution minimizes energy. As frequency decreases and crosses $\omega_{cr}$, the condensate localizes and the bright soliton appears. Therefore, $\omega_{cr}$ turns out to be the branching point at which $E_{b.s.}(\omega_{cr}) = E_{b.c.}(\omega_{cr})$ and $Q_{b.s.}(\omega_{cr}) = Q_{b.c.}(\omega_{cr})$. The reason behind the emergence of this point is exactly the stability of the classical BEC solution.

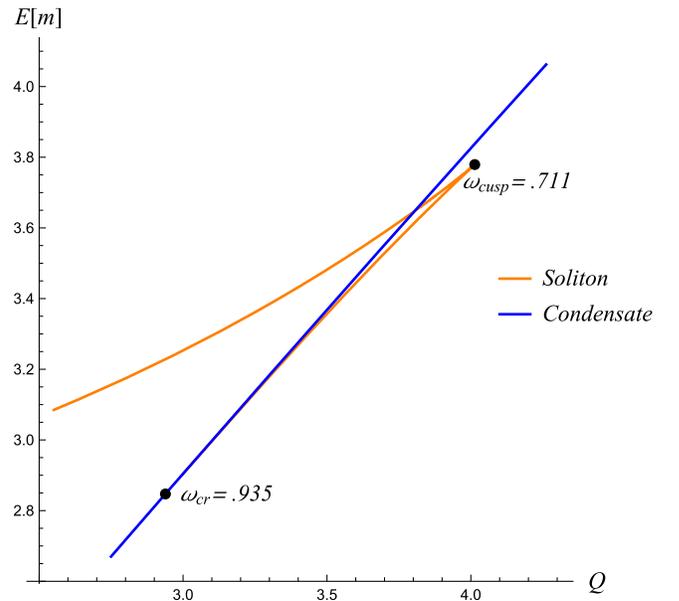

FIG. 2. Energy versus charge behavior of condensate and soliton. Here $L = 4\pi m^{-1}$ and $\lambda = 2m^2$.

Since the potential for this model is unbounded from below,[9] it is natural to question the classical stability of the system. This can be done by means of expanding fields in small perturbations around a given solution of classical equations, namely

$$\begin{cases} \phi(x,t) = \phi_{cl}(x,t) + \chi(x,t) \\ \phi^*(x,t) = \phi_{cl}^*(x,t) + \chi^*(x,t) \end{cases}, \quad (44)$$

where $\chi(x,t) = \mathcal{O}(\lambda^0)$ is the small perturbation, compared to $\phi_{cl} = \mathcal{O}(\lambda^{-1/2})$.

Next, we linearize equations of motions with respect to $\chi$ and $\chi^*$ and study the stability of $\chi(x,t)$ modes. In case of instability, we refer to the frequency of the exponentially growing mode as Lyapunov exponent.

Frequencies of modes comprising classical perturbations $\chi$ and $\chi^*$ are given by (see [10])

$$\gamma_+(p_n) = \sqrt{p_n^2 + 3\omega^2 - m^2 + \sqrt{m^4 - 6m^2\omega^2 + 4p_n^2\omega^2 + 9\omega^4}}, \quad (45)$$

---

[9]This plays no role in our analysis, because the field amplitude is lower than $2m/\sqrt{\lambda}$ and we do not enter the domain when boundlessness cannot be ignored anymore. In the general case, of course the theory can be made bounded by means of inclusion of higher order terms.

$$\gamma_-(p_n) = \sqrt{p_n^2 + 3\omega^2 - m^2 - \sqrt{m^4 - 6m^2\omega^2 + 4p_n^2\omega^2 + 9\omega^4}}, \tag{46}$$

where

$$p_n = \frac{2\pi n}{L}, \qquad n \in \mathbb{Z}. \tag{47}$$

Note that for solution (36), one perturbation mode is classically gapless i.e., $\gamma_-(0) = 0$. Therefore, in order to analyze the instability we focus on the nonzero modes. In particular, from (46) we see that the first mode which becomes imaginary, as $\omega$ decreases, is $p_1$ and it happens when $\omega$ becomes less than $\omega_{cr}$

$$\omega < \omega_{cr} = \sqrt{m^2 - \frac{2\pi^2}{L^2}}. \tag{48}$$

We therefore conclude that the condensate solution is classically stable for $\omega \in [\omega_{cr}, m)$ and unstable otherwise.[10]

In view of this analysis the meaning of the branching point is clear. In fact, the frequency at which the solitonic solution appears in the spectrum is exactly the same frequency (41) for which the first momentum mode (and therefore the condensate) becomes unstable. It follows that at $\omega_{cr}$ the spectrum splits into two branches of classical solutions for smaller frequencies (bigger charges): stable ones, namely bright solitons, and unstable ones, BEC's. We refer to the imaginary part of this frequency $\text{Im}(\gamma_-(p_1))$ as Lyapunov exponent, because it shows the rate of the exponential growth of this mode.

To sum up, we have the following situation: for $\omega \in [\omega_{cr}, m)$ the condensate solution is classically stable and unique for Eq. (35). In this region, the collective coupling, proportional to $(\lambda/m^2)f^2$, increases as $\omega$ decreases. Upon reaching $\omega_{cr}$ a phase transition takes place. In this sense, the collective coupling is strong enough as to allow for a localization of the solution. Correspondingly, the fluctuations on top of the condensate display at least an unstable mode indicating the fact that, for a given fixed charge, there exists a classical configuration with lower energy [as it can be seen by comparing (37) and (42)].

The above statement can easily be deduced from Fig. 2. As it is possible to see, the soliton trajectory emerges from the branching point $\omega_{cr}$. Moreover, as $\omega$ decreases (and the collective coupling increases), the soliton configuration localizes more and more up to the point where it becomes classically unstable (contrary to the classical theory). This happens in correspondence of $\omega_{cusp}$ which is the value for which the known instability condition is fulfilled [21,22]:

$$\left.\frac{dQ_{cl}}{d\omega}\right|_{\omega=\omega_{cusp}} < 0. \tag{49}$$

For lower $\omega$, the soliton solution corresponds to the points of the upper branch of Fig. 2.

Before moving forward it is important mentioning that although the potential is unbounded from below for the condensated (soliton), one can easily see from (36) [(39)], that the only source of classical instability under small perturbations is given by (41) [(49)]. Moreover, notice that tunneling phenomena are not relevant in our quantum study as they are exponentially suppressed, while we will see that quantum breaking happens exponentially fast.

### D. Saddle solution as initial condition

Appropriate initial conditions to numerically simulate (29) are necessary. Our goal is clear: we wish to study how the classical unstable condensate evolves as quantum fluctuations are dynamically taken into account. Therefore, a natural choice is to consider, at $t = 0$, the classical condensate solution. That means that, initially, the field $\phi$ is fixed by the stationarity condition

$$\frac{\delta S[\phi]}{\delta \phi_a} = 0 \tag{50}$$

while, for the propagator $G$, we need to invert the following operator

$$G_{ab}^{-1}(x,y) = i\left(\left(\partial^2 + m^2 + \frac{\lambda}{4}\phi_a^2\right)\delta_{ab} + \frac{\lambda}{2}\phi_a\phi_b\right) \times \delta^{(2)}(x-y). \tag{51}$$

This is already done in [10], and leads to

$$\tilde{G}_{ab}(t-\tau, x-y)$$
$$= \frac{1}{L}\int \frac{d\gamma}{2\pi}\sum_{n=-\infty}^{+\infty} e^{-i\gamma(t-\tau)+ip_n(x-y)}\tilde{G}_{ab}(\gamma, p_n)|_{p_n=\frac{2\pi n}{L}},$$
$$\tilde{G}_{ab}(\gamma, p_n)$$
$$= \frac{i((\omega^2 + \gamma^2 - p_n^2 - m^2 - \frac{\lambda}{4}f_d^2)\delta_{ac} + 2i\omega\gamma\epsilon_{ac} - \frac{\lambda}{2}f_af_c)}{(\gamma^2 - \gamma_+^2(p_n) + i0)(\gamma^2 - \gamma_-^2(p_n) + i0)}, \tag{52}$$

where the relation between $G$, $\tilde{G}$, and $\phi$, $f$ is given by

$$\begin{cases} \phi_a(t,x) = R_{ab}(\omega t)f_b \\ G_{ab}(t,x;\tau,y) = R_{ac}(\omega t)R_{bd}(\omega \tau)\tilde{G}_{cd}(t-\tau, x-y) \end{cases} \tag{53}$$

with $R_{ab}(\theta) \in SO(2)$ the standard rotational matrix, $f$ amplitude fixed by condition (36) and $\gamma_+(p)$ and $\gamma_-(p)$

---

[10]This situation reflects the presence of a Jeans instability and as the size gets bigger more modes become unstable. It occurs successively, namely, a given mode $p_n$ becomes unstable at $\omega_{cr}(p_n) = \sqrt{m^2 - 2\pi^2 n^2/L^2}$. Therefore we deduce that for fixed size $L$ there will be $\lfloor mL/\sqrt{2}\pi \rfloor$ unstable modes.

are defined in (45) and (46). Since for the tree level solution $\gamma_-(0)$ is gapless and $\gamma_-(p_1)$ has imaginary part, in order to specify initial conditions for $F$ from (52), at $t = 0$, the zeroth and first momentum modes of the statistical propagator were removed. This corresponds to a little shift away from the saddle point, anyway without affecting the dynamics in a relevant way. The mapping of the initial conditions between $G$ and $F$ and $\rho$ is explicitly derived in [10]. Let us remind here again that all initial conditions are derived with periodic boundary conditions, which are preserved during evolution due to translational invariance.

## V. NUMERICAL ANALYSIS

In the following we discuss our numerical results.

Equations (29) have been solved numerically using the Crank-Nicolson finite difference scheme for derivatives, and a trapezoidal rule for memory integrals. Moreover, since we are interested in the region where the classical spectrum is splitted between the condensate and the bright soliton, we confine our study to the range $\omega \in (\omega_{cusp}, \omega_{cr})$.[11] In this region, our choice of initial conditions, namely the unstable condensate solution, leads to the presence of one imaginary mode. As $L$ is increased, more and more modes display such behavior in the above-mentioned $\omega$ range, as it can be easily seen from (46). Therefore, the size of the box is purely dictated by reasons of simplicity: first of all the box size should be much bigger than the constituents Compton wavelength $\omega^{-1}$ and, second, the numerical analysis simplifies a lot if only one instability mode shows up in the explored $\omega$'s range. In view of this, we chose the box size to be $L = 4\pi m^{-1}$, for which $\omega_{cusp} \simeq 0.71m$, $\omega_{cr} \simeq 0.94m$ and only one instability mode is present, namely the first one $\gamma_-(p_1)$. In all the simulations we fixed $m = 1$. The general behavior of the quantum breaking is displayed for different $\omega$'s in Fig. 3 for strong and weak coupling. The behavior is very reminiscent to the one observed for the quantum breaking in the repulsive case for a stable condensate [10], although much faster. However, even if not shown explicitly, we notice the following feature: for all the simulations in the strong coupling regime, the total charge (19) varies on the displayed timescales by approximately the 1%, i.e., $Q(m^{-1})/Q(0) \sim 0.99$. This is comparable to the change in classical charge displayed in Fig. 3(a). However, due to the coupling being strong, we do not expect the resulting simulation to approximate appropriately the true quantum evolution anyway. The situation is qualitative different in the weak coupling regime Fig. 3(b). There, even though the coupling is weak, as $\omega$ is decreased (and the collective coupling is increased), there is a violation of the conservation of the total charge. This is, however, practically negligible (at least 100 times smaller and slightly growing with the collective coupling). This is no longer the case as we look at the simulation for longer times. For example, when $Q_{cl}(t) \approx 0.5 Q_{cl}(0)$ we generically notice an increasing violation of the total charge conservation. This underlies a failure of our numerical scheme as the exponential growth of $Q_q$ is so fast that it can no longer be well approximated after certain timescales. We now proceed with the discussion of the two main results of this work.

### A. Evolution along the BEC trajectory

The first result we wish to comment upon is the evolution of the unstable condensate.

We will introduce here two notions for the energy as we did for the charge. So, in full analogy with $Q_{cl}(t)$ and $Q_q(t)$ defined in (21) and (23) we define

$$E_{cl}(x^0) = \int_0^L dx^1 \left(\frac{1}{2}(\partial_{x^0}\phi_a)^2 + \frac{1}{2}(\partial_{x^1}\phi_a)^2 + V(\phi_a)\right) \quad (54)$$

and quantum part which we are not interested in writing explicitly here.

The sum of two is a conserved quantity

$$\frac{d(E_{cl}(t) + E_q(t))}{dt} = 0. \quad (55)$$

During the evolution both classical charge and energy diminish while full integrals of motion are conserved. In Fig. 4, the evolution of this classical quantities is shown for different $\omega$'s (different colors) and compared with respect to the branch of classical condensate solutions (blue line).

The points in the plot are numerical evaluations of the classical energy (54) and charge (38) at different times. Different colors there correspond to different initial configurations, namely different initial $\omega$. Therefore looking the dots of particular color, which coordinates at the plot are $(Q_{cl}(t), E_{cl}(t))$, one can see how the classical energy and classical charge of this configuration evolve with time. In Fig. 4(a) one can see how the numerical simulations fully evolve along the classical condensate trajectory. The reader might wonder why no deviations are seen, although lower energy configurations (at similar charge), namely the soliton (orange line), are present. The reason is that the bright soliton is a localized configuration, and therefore, due to the homogeneity of our initial conditions, the system in principle can only evolve into a superposition of such solutions. We must admit that here we actually do not know how to understand whether initial homogeneous configuration is tending to approach a superposition of solitons or not but there is a clear qualitative picture explaining the condensate evolution along the BEC $E_{cl}(Q_{cl})$ line.

---

[11]To avoid confusion we have to mention that for fixed $\omega$, the charges of soliton and condensate are different and they have, correspondingly, different energies [as it can be seen from (38) and (43)], though they coincide at the critical point $Q_{b.s.}(\omega_{cr}) = Q_{b.c.}(\omega_{cr})$.

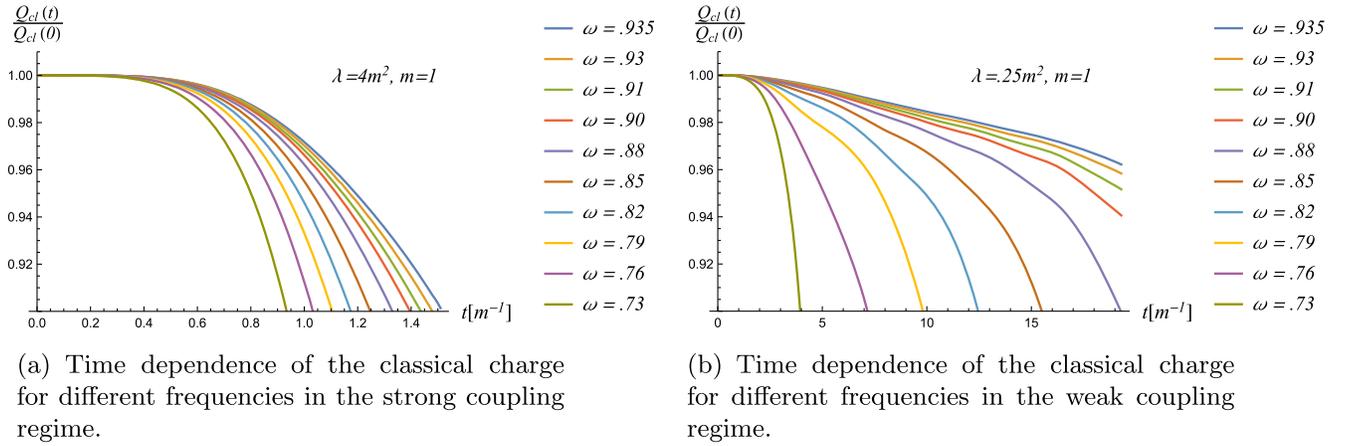

(a) Time dependence of the classical charge for different frequencies in the strong coupling regime.

(b) Time dependence of the classical charge for different frequencies in the weak coupling regime.

FIG. 3. Time evolution of classical charges.

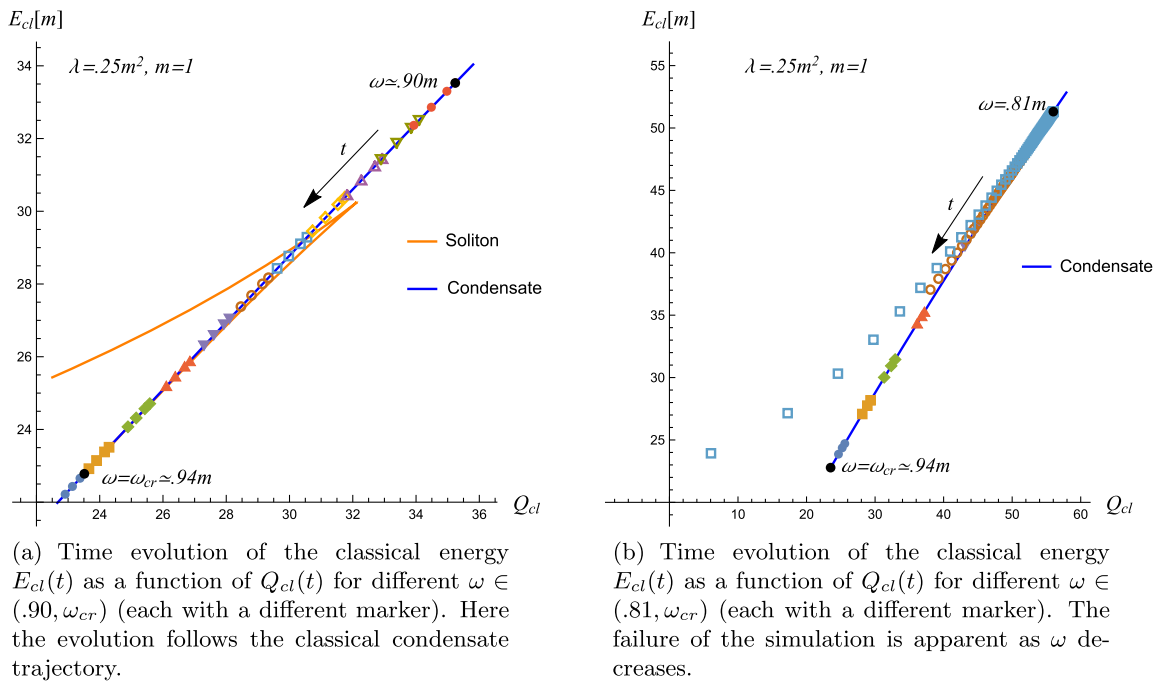

(a) Time evolution of the classical energy $E_{cl}(t)$ as a function of $Q_{cl}(t)$ for different $\omega \in (.90, \omega_{cr})$ (each with a different marker). Here the evolution follows the classical condensate trajectory.

(b) Time evolution of the classical energy $E_{cl}(t)$ as a function of $Q_{cl}(t)$ for different $\omega \in (.81, \omega_{cr})$ (each with a different marker). The failure of the simulation is apparent as $\omega$ decreases.

FIG. 4. Time evolution of the classical energy as a function of $Q_{cl}$ for different frequencies.

In total analogy with the Q-balls case, one can deduce the following relation between classical energy and charge of the BEC

$$\frac{dE_{cl}}{d\omega} = \omega \frac{dQ_{cl}}{d\omega}. \tag{56}$$

[12]Therefore, one can conclude that if classical charge is getting changed by some small amount $\delta Q_{cl}$, which in our case the part carried out by fluctuations, then the next homogeneous configuration stationarizing the local part of the energy is given by

$$E_{cl}(\delta t) \simeq E_{cl}(0) + \omega \delta Q_{cl}(\delta t), \tag{57}$$

which means that during quantum evolution, the preferable directions in the phase space is the one stationarizing the classical part as long as classical quantities remain dynamically dominant. This is true as long as $Q_{cl} \gg Q_q$, but in the end, upon reaching $t_{qb}$, this approximate relation does not need to hold, as well as our $\hbar$ perturbative expansion (See discussion in Sec. I), and evolution might deviate from this curve. Anyway, we will see in the next paragraph how the presence of a lower energy configuration affects the speed at which the system rolls along the BEC $E_{cl}(Q_{cl})$ trajectory.

---

[12]One can see once again from this relation the role played by $\hbar$ in the analysis. Namely, energy and $\omega$ are classical quantities, finite in the $\hbar \to 0$ limit. Therefore, restoring units one explicitly sees $[E\omega^{-1}] = [\hbar] = [Q_{cl}]$. Hence, the mean-field value of the dimensionless operator $\langle \hat{Q} \rangle$ is connected to $\hbar$ in the following way $Q_{cl} = \hbar \langle \hat{Q} \rangle$ and for $\hbar \to 0 \langle \hat{Q} \rangle \to \infty$, $Q_{cl} =$ finite.

The situation is different in Fig. 4(b) where the behavior of simulations is shown for lower $\omega$'s. As one can see, some of the simulations (blue squares) start deviating from the condensatelike evolution. Such deviations increase as $\omega$ decreases. Such phenomenon is analogue to the one observed for the strong coupling case. Namely, the exponential growth becomes so fast, as to lead to a failure of the numerical scheme. In fact, correspondingly with these deviations, non-negligible—yet very small—violations of total charge emerge. However, it should be noted that for small timescales, the evolution of these simulations is still reliable.

### B. Quantum break-time

In order to capture the rate at which the system evolves along the condensate trajectory (cf. Fig. 4), we use criterion (26) to obtain $t_{qb}$. Moreover, because of the above mentioned numerical issues, we fix quantum breaking as the time such that $Q_q(t_{qb}) = 0.1 Q_{cl}(t_{qb})$. In this way, on the obtained quantum breaking timescales, most of the simulations have a very well conserved charge (within .1% deviations). The dependence of $t_{qb}$ on $\log Q / \text{Im}(\gamma_-(p_1))$ is explicitly shown in Fig. 5, where the quantum breaking time is shown with respect to $\omega$. One can see how in this coupling region, i.e., $\lambda \leq 0.1 m^2$, the breaking time is indeed captured by the relation

$$t_{qb} \simeq \frac{\log Q}{\text{Im}(\gamma_-(p_1))} + \text{constant} \tag{58}$$

where $\text{Im}(\gamma_-(p_1))$ is the Lyapunov exponent, therefore confirming (2) mentioned in the introduction. Let us discuss this relation and the corresponding physics in more details.

First, it is very important to stress that such a logarithmic behavior is controlled by the Lyapunov exponent which sets the system instability. As long as a homogeneous solution exhibits this form of instability, the leading initial time behavior is apparently governed by this mode arising in the Green's function. Therefore we can roughly say that for $\delta t \sim 1/\text{Im}(\gamma_-(p_1))$

$$Q_{cl}(\delta t) \simeq Q_{cl}(0) + A(1 - e^{\text{Im}(\gamma_-(p_1))\delta t}) \tag{59}$$

and we can infer from this, that using our quantum breaking criterion $Q_{cl}(t_{qb})/Q_{cl}(0) \simeq r$, we obtain (58). Indeed, the aforementioned constant in (58) depends only on the chosen criterion $r$ and is almost independent on any parameters of the model.

From the quantum point of view, this law means that to undergo quantum breaking, the system has to wait for a significant amount of quanta to decay, which leads to logarithmic dependence of quantum break-time on charge. This mechanism is in contrast with the classical picture, where departure from stationary solution is set by inhomogeneities introduced by means of initial classical perturbations. To clarify this difference let us briefly recap how the decay due to classical perturbations takes place. In fact, for this to happen, the inhomogeneous perturbation spectrum needs to include an unstable mode, which, in the system under consideration, corresponds with the first momentum mode. In turn, the presence of such a mode leads to the development of inhomogeneities in the system on a time scale fixed by the Lyapunov exponent as $1/\text{Im}(\gamma_-(p_1))$, leading to the localization of the soliton. Thus, we see that departure occurs almost immediately, independently of the initial configuration itself (namely amount of quanta realizing the initial state). This kind of transitions were clearly demonstrated for unstable Q-balls (Q-clouds) as they are perturbed and, after a timescale set by the instability, they move to a stable configuration, dropping some charge along the process [23]. In addition, we must stress again, that not all the perturbations can destroy BEC classically, but only those which contain the unstable mode. This is totally different from what happened in our case as the evolution is driven by quantum effects preserves translational invariance. In fact, we saw in Fig. 4(a) classical quantities evolve along the classical homogeneous BEC trajectory, although the growth of the unstable mode within the Green's function being present (note that at $t = 0$ we input this unstable mode to be zero). Therefore, we see that quantum and classical process are significantly different. In particular, the quantum decay is more general as it takes place independently whether or not homogeneity is preserved by initial conditions.

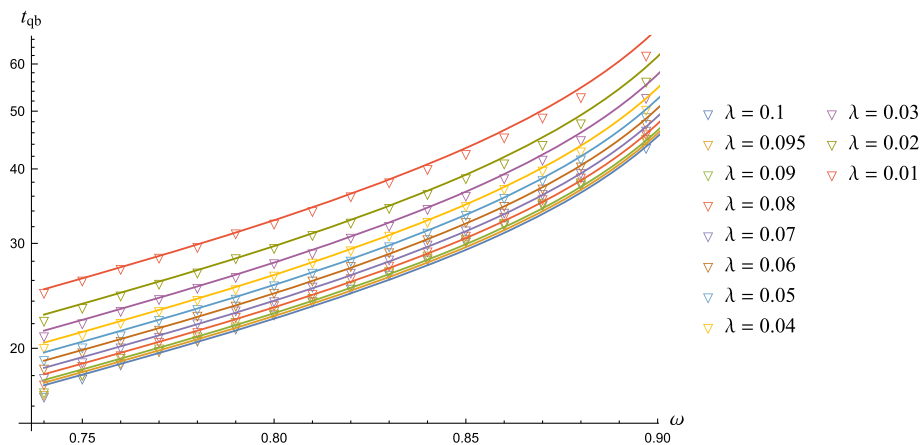

FIG. 5. Quantum breaking time dependence on $\omega$ compared to analytical estimation. Solid lines are functions $\log Q(\omega, \lambda)/\text{Im}(\gamma_-(p_1))$ for different couplings and triangles are quantum breaking times extracted from simulations.

One clear feature of Fig. 5 is the deviation from the logarithmic scaling as the coupling is increased. We believe this to be due to the above discussed failure of the numerical scheme.

## VI. CONCLUSIONS AND OUTLOOK

In this work we studied the behavior of a relativistic BEC near its classical instability. It was found that the quantum break-time, namely the moment when departure from the classical solution become significant, scales logarithmically with dimensionless charge $Q$ and is controlled by the Lyapunov exponent $\gamma$ characterizing the classically instability of the system, namely

$$t_{qb} = \gamma^{-1} \log Q + C, \quad (60)$$

with $C$ a small constant caused by the way we extracted numerically $t_{qb}$. This result for quantum break-time is similar to the one derived in [7] [see Eq. (2)]. However, being our study in the relativistic regime, we see that the number of constituents is replaced with charge $N \to Q$ [compare (60) with (1)]. These two results need not to be similar as the spectrum of the relativistic model is much broader. However, as it follows from our analysis, (60) is indeed the natural extension of the result of [7] in the relativistic regime.

To obtain (60), we worked within the 2-PI formalism, keeping the second order in loop-expansion, which allowed us to numerically simulate the Minkowski time evolution of the system. After setting as initial conditions the tree level stationary condensate solution, we introduced a new general criterion to measure its deviation from the classical solution, namely we chose as observable of interest the ratio between $Q_{cl}$ which is the functional of 1-point expectation value and $Q_q$ given by connected part of the propagator evolved in time. A natural criterion to obtain the quantum break time, therefore, relies on the dynamical condition $Q_{cl} \approx Q_q$ (see (26). Such timescale, is what we refer to as $t_{qb}$ in (60). This condition, already introduced by us in [10], turns out to be very simple as it requires the measurement of only an integral quantity, to be compared with [7], where $t_{qb}$ was calculated via entanglement arguments. It is therefore quite intriguing that such different methods do agree. Note that such condition might be imposed for any other integral quantity (e.g., energy) as long as its conservation constrains the dynamical evolution of the system. Therefore the criterion introduced in [10] proves general and with a wide applicability range.

Another interesting result of this article is related to the breaking trajectory of the condensate. In fact, the evolution takes place along the branch of classical solution $E_{cl}(Q_{cl})$ as long as charge is steadily getting carried away by quantum fluctuations according to (56) and (57) (see Fig. 4). There, the classical part of the energy with respect to classical charge evolves along the branch of classical solutions given by (37) and (38). This is due to the fact that homogeneity is preserved in our equations. In fact, as discussed above, there is a preferable direction in the phase space of classical configurations set by (56). Indeed, as soon as quantum breaking takes place, we do not expect this kind of relation to hold and the system might evolve differently from there onward. Further investigation of this issue, although interesting, is beyond the scope of this work. Moreover it can be seen that as either the coupling or the collective coupling is increased, deviations from such trajectory are observed in Fig. 4. We believe this to be due to a failure of our numerical scheme. We therefore excluded such simulations when studying the functional dependence of the quantum break time.

## ACKNOWLEDGMENTS

First of all we would like to thank IMPRS research school of Max-Planck Institute for Physics for giving us the opportunity to participate in their program for Ph.D. students and Arnold Sommerfeld center for having great scientific environment there. As well we are grateful to our supervisor Gia Dvali for inspiring and useful discussions and for reading the manuscript. A. K. wants to thank Emin Nugaev for discussing the results and for useful advises. M. Z. is thankful to Lorentz center in Leiden for its kind hospitality and fruitful discussions he had there while working on the project. Finally both the authors thank the anonymous referee whose comments helped to significantly improve the manuscript.

[1] G. Dvali and C. Gomez, Black hole's quantum N-portrait, Fortschr. Phys. **61**, 742 (2013).
[2] G. Dvali and C. Gomez, Black hole macro-quantumness, arXiv:1212.0765.
[3] G. Dvali and C. Gomez, Black hole's 1/n hair, Phys. Lett. B **719**, 419 (2013).
[4] G. Dvali and C. Gomez, Quantum compositeness of gravity: Black holes, AdS and inflation, J. Cosmol. Astropart. Phys. 01 (2014) 023.
[5] G. Dvali, C. Gómez, and S. Zell, Quantum break-time of de sitter, J. Cosmol. Astropart. Phys. 06 (2017) 028.
[6] G. Dvali and S. Zell, Classicality and quantum break-time for cosmic axions, J. Cosmol. Astropart. Phys. 07 (2018) 064.
[7] G. Dvali, D. Flassig, C. Gomez, A. Pritzel, and N. Wintergerst, Scrambling in the black hole portrait, Phys. Rev. D **88**, 124041 (2013).
[8] P. Hayden and J. Preskill, Black holes as mirrors: Quantum information in random subsystems, J. High Energy Phys. 09 (2007) 120.
[9] J. Berges, S. Borsányi, U. Reinosa, and J. Serreau, Nonperturbative renormalization for 2pi effective action techniques, Ann. Phys. (Amsterdam) **320**, 344 (2005).
[10] A. Kovtun and M. Zantedeschi, Breaking BEC, J. High Energy Phys. 07 (2020) 212.
[11] J. Berges, S. Borsanyi, U. Reinosa, and J. Serreau, Renormalized thermodynamics from the 2PI effective action, Phys. Rev. D **71**, 105004 (2005).
[12] J. Berges, Controlled nonperturbative dynamics of quantum fields out-of-equilibrium, Nucl. Phys. **A699**, 847 (2002).
[13] C. de Dominicis and P. C. Martin, Stationary entropy principle and renormalization in normal and superfluid systems. I. Algebraic formulation, J. Math. Phys. (N.Y.) **5**, 14 (1964).
[14] C. De Dominicis and P. C. Martin, Stationary entropy principle and renormalization in normal and superfluid systems. II. Diagrammatic formulation, J. Math. Phys. (N.Y.) **5**, 31 (1964).
[15] J. M. Cornwall, R. Jackiw, and E. Tomboulis, Effective action for composite operators, Phys. Rev. D **10**, 2428 (1974).
[16] J. Berges, Introduction to nonequilibrium quantum field theory, AIP Conf. Proc. **739**, 3 (2004).
[17] H. van Hees and J. Knoll, Renormalization in self-consistent approximation schemes at finite temperature. 3. Global symmetries, Phys. Rev. D **66**, 025028 (2002).
[18] J. Berges, Introduction to nonequilibrium quantum field theory, in *AIP Conference Proceedings* (American Institute of Physics, New York, 2004), Vol. 739, pp. 3–62.
[19] L. D. Carr, C. W. Clark, and W. P. Reinhardt, Stationary solutions of the one-dimensional nonlinear schrödinger equation. II. Case of attractive nonlinearity, Phys. Rev. A **62**, 063611 (2000).
[20] R. Kanamoto, H. Saito, and M. Ueda, Quantum phase transition in one-dimensional bose-einstein condensates with attractive interactions, Phys. Rev. A **67**, 013608 (2003).
[21] N. Vakhitov and A. A. Kolokolov, Stationary solutions of the wave equation in the medium with nonlinearity saturation, Radiophys. Quantum Electron. **16**, 783 (1973).
[22] R. Friedberg, T. Lee, and A. Sirlin, Class of scalar-field soliton solutions in three space dimensions, Phys. Rev. D **13**, 2739 (1976).
[23] A. Panin and M. Smolyakov, Classical behaviour of Q-balls in the Wick–Cutkosky model, Eur. Phys. J. C **79**, 150 (2019).